\documentclass[aps,pra,twocolumn,groupedaddress,showkeys]{revtex4}
\usepackage{graphicx}
\usepackage{dcolumn}
\usepackage{bm}
\usepackage{amsmath}
\usepackage{amssymb}
\usepackage{color}
\usepackage{slashed}
\usepackage{hyperref}
\usepackage{tikz}
\usepackage{pgfplots}
\usepackage{array}
\pgfplotsset{width=7.5cm}
\newcommand{\qbar}{\raisebox{-1.5ex}[\height][0pt]{$\mathchar'26$}\mkern-9mu q}
\newcommand{\boldqbar}{\raisebox{-1.45ex}[\height][0pt]{$\mathchar'26$}\mkern-12.5mu \mathbf{q}}

\begin{document}
\begin{flushleft}
\texttt{RBI-ThPhys-2023-39}\\
\texttt{ZTF-EP-23-05}
\end{flushleft}

\title{Gravitational probe of $\,\boldqbar$uantum spacetime}

\author{Nikola Herceg}
\email{nherceg@irb.hr}

\author{Tajron Juri\'c}
\email{tjuric@irb.hr}

\author{Andjelo Samsarov}
\email{asamsarov@irb.hr}
\affiliation{Rudjer Bo\v{s}kovi\'c Institute, Bijeni\v cka  c.54, HR-10002 Zagreb, Croatia}

\author{Ivica Smoli\'c}
\email{ismolic@phy.hr}
\affiliation{Department of Physics, Faculty of Science, University of Zagreb, 10000 Zagreb, Croatia}

\author{Kumar S. Gupta}
\email{kumars.gupta@saha.ac.in}
\affiliation{Theory Division, Saha Institute of Nuclear Physics, 1/AF Bidhannagar, Kolkata 700064, India}

\date{\today}

\begin{abstract}
A quest for phenomenological footprints of quantum gravity is among the central scientific tasks in the rising era of gravitational wave astronomy. We study gravitational wave dynamics within the noncommutative geometry framework, based on a Drinfeld twist and newly proposed noncommutative Einstein equation, and obtain the leading quantum correction to Regge-Wheeler potential up to first order in the noncommutativity parameter. By calculating the quasinormal mode frequencies we show that the noncommutative Schwarzschild black hole remains stable under axial gravitational perturbations.
\end{abstract}


\maketitle
\section{Introduction} 

The discovery of gravitational waves \cite{LIGOScientific:2016aoc, LIGOScientific:2016sjg,  LIGOScientific:2017vwq,  nature} has ushered in a new era in astrophysics and cosmology. The findings so far have involved black holes or neutron stars which are governed by general relativity where the gravitational field is classical. Several studies have suggested the tantalizing possibility that gravitational waves may be able to encode signatures of certain quantum aspects of gravity, such as the effects of quantum fluctuations of the gravitational field on infalling bodies \cite{wilczek} and gravitational wave signatures of black hole area discretization \cite{pullin}. In addition, the future gravitational wave detectors such as LISA offer the possibility to test the effects of modified dispersion relation on the gravitational wave propagation \cite{lisa}. 

The aim of this paper is to study the effects of quantum structure of spacetime on gravitational wave formation. It is well known \cite{dfr1,dfr2} that classical general relativity and the quantum uncertainty principle lead to noncommutative (NC) geometry, where the smooth manifold structure of the spacetime is replaced by a NC algebra, which is a candidate for the quantum spacetime. The gravitational wave emission in the classical spacetime was first discussed by Regge and Wheeler \cite{Regge:1957td}, followed by a number of seminal papers \cite{Edelstein:1970sk,Zerilli:1970se, natureV, visv, Chandrasekhar:1975zza, Schutz:1985km, Thorne:1980ru}, which have been discussed in several reviews \cite{Kokkotas:1999bd, Konoplya:2011qq}. In these works, the quasi-normal modes (QNM) of the gravitational wave are described by a Schr\"{o}dinger-type equation, in which the potential carries the information of the underlying spacetime. As a first step of discussing gravitational wave propagation in quantum spacetime, we adapt the analysis of Regge and Wheeler \cite{Regge:1957td} to the NC framework, thereby obtaining the NC version of the Regge-Wheeler potential which we will refer to as $\qbar$-deformed or $\qbar$uantum Regge-Wheeler ($\qbar$RW) potential.

The symmetries of a spacetime are encoded in the algebra of infinitesimal diffeomorphisms, while the deformed symmetries of a NC spacetime can be encoded in the Hopf algebra of the deformed diffeomorphisms. This Hopf algebra framework has been shown to be useful in the area of NC quantum field theory \cite{Chaichian:2004yh, Chaichian:2000si, Balachandran:2010xk} and can be used to construct the NC differential geometry leading to NC generalizations of concepts such as Lie derivative, connection, covariant derivative, torsion and curvature tensor \cite{Aschieri:2005zs, Aschieri:2005yw, Aschieri:2009qh}. These geometric objects are crucial for our study of NC effects to gravity. The NC deformations are introduced via Drinfeld twist \cite{majid}. In this paper we will use the geometric framework developed in \cite{Aschieri:2005zs, Aschieri:2005yw, Aschieri:2009qh}, but with an alternative and more natural definition of NC Einstein manifolds. Other ways for grasping NC effects of gravity are presented in \cite{Chaichian:2007we, Chaichian:2007dr, Calmet:2005qm} where the  NC tetrad formalism was used, and in \cite{Kobakhidze:2007jn} where coupling the Einstein equation with a NC energy-momentum tensor was performed. Also, it is important to mention that our NC framework differs form \cite{Connes:1994yd, Chamseddine:1996zu, Chamseddine:1996rw} where the derivations obey the standard Leibniz rule, while in our paper we have a deformed Leibniz rule and $\mathcal{R}$-symmetry governing the whole construction of NC differential geometry.

Our analysis within the NC framework leads to several key results. First, the $\qbar$uantum correction to the Regge-Wheeler equation appears at the first order of the NC deformation parameter. This is in marked contrast to most results in the literature, where such corrections appear only in the second order \cite{Chaichian:2007we, Chaichian:2007dr, Calmet:2005qm, Harikumar:2006xf}. Our results are thus much better suited for confronting with empirical data. The second important result has to do with stability of black holes in the quantum spacetime. As we explicitly demonstrate,  the Schwarzschild black hole is stable under linear perturbations in the NC framework. This may have implications for the stability of black holes at the Planck scale, where the NC effects are expected to be dominant. Finally, in the limit of the vanishing NC parameter, we smoothly recover the standard results obtained within the commutative framework. 

The paper is structured as follows. After introducing the relevant geometric objects from the NC differential geometry framework, we postulate an equation for NC Einstein manifolds. We study this equation in the fixed background of a Schwarzschild black hole with the axial gravitational perturbations. All the calculations are performed up to the first order in the NC parameter $\qbar$ and gravitational perturbation $h$. Finally, we obtain a Schr\"{o}dinger like equation that is governed by the $\qbar$-deformed Regge-Wheeler potential. This way we generalize the work of Regge and Wheeler \cite{Regge:1957td} to the NC framework. We discuss some features of the $\qbar$-deformed Regge-Wheeler potential and $\qbar$-QNM's.

\section{NC differential geometry}
\label{NCDG}
In order to find gravitational perturbations for the Schwarzschild metric one needs to solve the linearized vacuum Einstein equation
\begin{equation}\label{a}
R_{\mu\nu}(g)=0,
\end{equation}
where $R_{\mu\nu}$ is the Ricci tensor that depends on the metric $g=g_s+h$ and $g_s$ is the Schwarzschild background metric, while $h$ is the perturbation in which we expand up to linear order. In order to find  a NC analog of \eqref{a} we use the NC differential geometry framework \cite{Aschieri:2005zs, Aschieri:2005yw, Aschieri:2009qh} for the Moyal type deformations given by the twist
\begin{equation}\label{twist}
	\mathcal{F}=e^{- i \Theta^{\alpha\beta}\partial_{\alpha}\otimes\partial_{\beta}}=f^{A}\otimes f_{A},
\end{equation}
with the inverse $\mathcal{F}^{-1} = \bar{f}^A \otimes \bar{f}_A$, where $\Theta^{\alpha\beta}$ is a constant antisymmetric matrix and $\left\{\partial_{\mu}\right\}$ are coordinate basis vector fields. This twist generates a NC space defined by the algebra
\begin{equation}
[x^{\mu}\stackrel{\star}{,}x^{\nu}]=x^{\mu}\star x^{\nu}-x^{\nu}\star x^{\mu}=i \Theta^{\mu\nu},
\end{equation}
where the NC $\star$-product is given by
\begin{equation}
f\star g=fg+i\Theta^{\alpha\beta}\frac{\partial f}{\partial x^{\alpha}}\frac{\partial g}{\partial x^{\beta}}+\mathcal{O}(\Theta^2), \quad \forall \ f,g\in\mathcal{C}^{\infty}.
\end{equation}
This NC $\star$-algebra $(\mathcal{C}^{\infty}, \star)$ is a generalization of the algebra of smooth functions on a manifold with point-wise multiplication. The symmetries of this algebra are encoded in the $\star$-Lie algebra of infinitesimal diffeomorphisms and the corresponding Hopf algebra \cite{Aschieri:2005zs, Aschieri:2005yw, Aschieri:2009qh, Juric:2022bnm}. The most important feature of this framework is that the usual Leibniz rule of how symmetry generators (vector fields) act on the NC algebra is twisted. This leads to twisted notions of Lie derivative $\pounds^{\star}$, connection $\hat{\nabla}$, curvature tensor $\hat{R}$ and torsion $\hat{T}$ (for more details see \cite{Aschieri:2005zs, Aschieri:2005yw, Aschieri:2009qh, Juric:2022bnm, novi}). The NC curvature tensor $\hat{R}$ is given by
\begin{equation}
	\hat{R}(u,v,w)=\hat{\nabla}_{u}\hat{\nabla}_{v}w - \hat{\nabla}_{\bar{R}^A(v)}\hat{\nabla}_{\bar{R}_A(u)} w - \hat{\nabla}_{[u,v]_{\star}}w,
\end{equation}
where $u$, $v$ and $w$ are vector fields, $[u,v]_{\star}=u\star v-\bar{R}^A(v) \star \bar{R}_{A}(u)$ is the $\mathcal{R}$-permuted commutator and $\bar{R}^A(v) \star \bar{R}_{A}(u)$ is the deformed action of the inverse $\mathcal{R}$-matrix $\mathcal{R}^{-1}=\bar{R}^A\otimes\bar{R}_{A}=\mathcal{F}\mathcal{F}^{-1}_{21}=f^{A}\bar{f}_{B}\otimes f_{A}\bar{f}^{B}$. The components of the NC Ricci tensor $\hat{R}_{\mu\nu}$ are given by  evaluating $\hat{R}$ on basis vector fields $\left\{\partial_{\mu}\right\}$ and contracting with basis dual one forms $\left\{dx^{\mu}\right\}$. For the case of a Moyal twist \eqref{twist} consisting of the basis vector fields, the NC Ricci is given by
\begin{equation}
\hat{R}_{\mu\nu}=\left\langle dx^{\alpha}, \hat{R}(\partial_{\alpha}, \partial_{\mu}, \partial_{\nu})\right\rangle_\star
\end{equation}
and the expressions for the components of NC curvature $\hat{R}$ and NC torsion $\hat{T}$ are analogous to their commutative counterparts \cite{Aschieri:2005zs, Aschieri:2005yw, Aschieri:2009qh, Juric:2022bnm}
\begin{equation}\begin{split}\label{RiT}
&\hat{R}^{\ \ \ \ \sigma}_{\mu\nu\rho}=\partial_{\mu}\Sigma_{\nu\rho}^{\ \ \ \sigma}-\partial_{\nu}\Sigma_{\mu\rho}^{\ \ \ \sigma}+\Sigma_{\nu\rho}^{\ \ \ \tau}\star\Sigma_{\mu\tau}^{\ \ \ \sigma}-\Sigma_{\mu\rho}^{\ \ \ \tau}\star\Sigma_{\nu\tau}^{\ \ \ \sigma},\\
&\hat{T}^{\ \ \ \rho}_{\mu\nu}=\Sigma_{\mu\nu}^{\ \ \ \rho}-\Sigma_{\nu\mu}^{\ \ \ \rho},
\end{split}\end{equation}
where the coefficients $\Sigma_{\mu\nu}^{\ \ \ \rho}$ are uniquely determined by the choice of connection
\begin{equation}
\hat{\nabla}_{\mu}\partial_{\nu}=\Sigma_{\mu\nu}^{\ \ \ \rho}\star\partial_\rho=\Sigma_{\mu\nu}^{\ \ \ \rho}\partial_\rho.
\end{equation}
In the first equality we are using the fact that $\hat{\nabla}_{\mu}$ is mapping $\star$-vector fields to $\star$-vector fields and any $\star$-vector field can be written as a linear combination of basis vectors $\Sigma_{\mu\nu}^{\ \ \ \rho}\star\partial_\rho$, while in the second equality we use the fact that the twist is Moyal and each slot acts as a Lie derivative, and since $\pounds_{\partial_{\mu}}(\partial_{\alpha})=[\partial_{\mu},\partial_{\alpha}]=0$ the $\star$-product reduces to the usual pointwise product. \\
There is a unique torsion free $\hat{T}=0$ and metric compatible $\hat{\nabla}_{\mu}g=0$ NC-Levi-Civita connection given by \cite{Aschieri:2009qh}
\begin{equation}\label{sLC}
\Sigma_{\mu\nu}^{\ \ \ \rho}=\frac{1}{2}g^{\star\rho\sigma}\star\left(\partial_{\mu}g_{\nu\sigma}+\partial_{\nu}g_{\mu\sigma}-\partial_{\sigma}g_{\mu\nu}\right),
\end{equation}
where $g^{\star\rho\sigma}$ is the unique $\star$-inverse satisfying $g^{\star\alpha\rho}\star g_{\rho\beta}=\delta^{\alpha}_{\beta}$ and $g_{\mu\rho}\star g^{\star\rho\nu}=\delta^{\nu}_{\mu}$
and is explicitly given as 
\begin{equation}
g^{\star\alpha\beta}=g^{\alpha\beta}-g^{\gamma\beta} i \Theta^{AB}(\partial_A g^{\alpha\sigma})(\partial_B g_{\sigma\gamma})+\mathcal{O}(\Theta^2).
\end{equation}
In order to calculate the $\star$-inverse, connection and curvature, we need to specify the exact form of the twist. 
While we can expect that the NC scale is around the Planck length, there is no general consensus on the exact algebra satisfied by the spacetime coordinates.
Noncommutativity is almost always investigated from local point of view by deforming the Poincar\'e algebra, or on a flat, maximally symmetric manifold. 
In both cases introducing NC structure usually breaks or alters the usual concept of symmetry.
However, in the Schwarzschild spacetime, global symmetry is already reduced --- it is spherical in space and static in time.
Therefore, it is reasonable to assume that effects of noncommutativity on macroscopic phenomena in vicinity of the black hole can be effectively described by a NC algebra that shares some of the symmetries of the black hole.
In such case we have Killing vector fields such as $\partial_t$ and $\partial_\phi$ at our disposal and we can use them to ``build'' our twist $\mathcal{F}$, for example we can write the so-called angular twist \cite{Ciric:2017rnf, Gupta:2022oel, Ciric:2018angular}
\begin{equation}\label{angtw}
\mathcal{F}=1\otimes 1 -\frac{ia}{2}(\partial_t \otimes\partial_\phi - \partial_\phi \otimes\partial_t)+\mathcal{O}(a^2), \quad a\in\mathbb{R}.
\end{equation}
For this type of twist we immediately have a ``no-go theorem'':
If the twist is built only out of Killing vector fields, then any commutative solution is also a NC solution \cite{Aschieri:2009qh}. 
This is due to the fact that $\pounds_{\partial_t}g_s=\pounds_{\partial_\phi}g_s=0$, so all the NC corrections in \eqref{RiT} vanish for the case of pure Schwarzschild metric $g_s$.

	For example, the $\star$-metric inverse is equal to the undeformed one since $g^s_{\mu \nu}\star g_s^{\nu \rho} = g^s_{\mu \nu} g_s^{\nu \rho} + i a/2 \left( \partial_t g^s_{\mu \nu} \partial_\phi g_s^{\nu \rho} \right) + O(a^2) = \delta_\mu^{~\rho} + 0$. Connection, curvature and vacuum Einstein equation remain undeformed for the same reason.

The same conclusion follows if the twist is semi-Killing, i.e. for the twist of the form
\begin{equation}\label{semi}
\mathcal{F}=1\otimes 1 -\frac{ia}{2}\left[(\alpha\partial_t +\beta\partial_\phi)\wedge v\right]+\mathcal{O}(a^2) \, ,
\end{equation}
with $v = v^\alpha(x) \partial_\alpha$.
This is probably the reason why there are not so many investigations around noncommutative corrections to gravity using this formalism (with the exception being \cite{Ohl:2009pv, Schenkel:2010sc}). However, since we will be interested in the perturbation of the Schwarzschild metric, i.e. $g=g_s+h$, we will not build our twist out of Killing vectors of the full metric $g$, but rather just out of Killing vectors of the Schwarzschild background. So, now twist \eqref{angtw} is a pseudo-Killing twist and since $\pounds_{\partial_t}g_s=\pounds_{\partial_\phi}g_s=0$ and $\pounds_{\partial_t}h\neq 0\neq\pounds_{\partial_\phi}h$, we see that the lowest non-vanishing NC corrections to \eqref{RiT} are quadratic in the perturbation $h$, namely the linear parts vanish. The conclusion is that if we want to respect the full symmetry of the background, then the NC effects are inherently \textit{nonlinear}, and to further study this we would have to go beyond the linearized  equation for $h$ even in the commutative case. This line of research, although interesting, we leave for future studies. Fortunately, we found that we could keep some of the symmetries by studying the so-called semi-pseudo Killing twists like \eqref{semi}. Now, for the twist \eqref{semi} the NC corrections to \eqref{RiT} that are linear in $h$ \textit{do exist}. For an extensive analysis of the aforementioned results in greater generality we refer the reader to \cite{novi}.
From now on in this paper, we will restrict our analysis for a special choice of the twist given by
\begin{equation}\label{qtwist}
\mathcal{F}=1\otimes 1 -\frac{i\qbar}{2}\partial_r\wedge\partial_\phi+\mathcal{O}(\qbar^2), \quad \qbar\in\mathbb{R}.
\end{equation}
and the ensuing theory and all observables obtained from it we will refer to as $\qbar$-deformed or just $\qbar$uantum. Also notice that the $\qbar$-twist \eqref{qtwist} leads to a special type of NC space, i.e $\qbar$-space, with the only nonvanishing $\star$-commutator being
\begin{equation}\label{rphialg}
[r\stackrel{\star}{,}\phi]=i\qbar,
\end{equation}
studied also in \cite{Kim:2007nx}. Algebra \eqref{rphialg} when transformed in the Cartesian coordinates is related to $\kappa$-deformed spaces \cite{C1, C2, L1, L2, L3, L4}.

\section{Quantum Regge-Wheeler}
\label{QRW}
There is still one more step left before we can derive the $\qbar$-deformed gravitational perturbations and $\qbar$uantum Regge-Wheeler potential. In the previous section we outlined the NC differential geometry and got the NC Ricci tensor $\hat{R}_{\mu\nu}$, but we still don't have  the NC version of \eqref{a}. Namely, a simple postulate such as $\hat{R}_{\mu\nu}=0$, which is compatible with the commutative limit, in general leads to an over-complete system of partial differential equations or a trivial solution that $\qbar = 0$ and therefore is not the right choice \footnote{This is actually an equation that determines the dynamics and should be derived from some action principle or using the $\star$-derived NC Bianchi identity in order to determine the possible divergentless tensors, namely NC analogs of the Einstein tensor $G_{\mu\nu}=R_{\mu\nu}-1/2g_{\mu\nu}R$. The proposal of NC Einstein tensor in \cite{Aschieri:2009qh} is not respecting the $R$-symmetry and leads to overcompleteness problems. We plan to report on the analysis of NC Einstein tensor elsewhere.}. The question arises why is this so. Well, $\hat{R}_{\mu\nu}$ is not a symmetric tensor by definition, so forcing $\hat{R}_{\mu\nu}=0=\hat{R}_{\nu\mu}$ is contradictory from the NC differential geometry point of view \footnote{In the commutative case $R_{\mu\nu}$ is symmetric by definition, so imposing it to be zero is not a problem.} \cite{novi}. 
Instead, the NC Ricci $\hat{R}_{\mu \nu}$ may be upgraded into a more general object ${\rm R}_{\mu \nu}$ that respects the $\mathcal{R}$-symmetry governed by the twist $\mathcal{F}$ \cite{novi}.
Therefore \textit{we postulate} the NC Einstein manifolds as those with the metric satisfying
\begin{equation}\label{NCein}
\hat{{\rm R}}_{\mu\nu}(g)=0,
\end{equation}
where $\hat{{\rm R}}_{\mu\nu}$ is the $\mathcal{R}$-symmetrized NC Ricci defined by 
\begin{equation}
\hat{{\rm R}}_{\mu\nu} = \frac{1}{2}\left\langle dx^{\alpha}, \hat{R}(\partial_{\alpha}, \partial_{\mu}, \partial_{\nu})+\hat{R}(\partial_{\alpha}, \bar{R}^{A}(\partial_\nu), \bar{R}_{A}(\partial_\mu)\right\rangle_\star.
\end{equation}
This is a generalization of vacuum equation proposed in \cite{Aschieri:2005zs, Aschieri:2009qh}.
In the case of $\qbar$-deformations \eqref{qtwist} the $\mathcal{R}$-symmetry reduces to usual symmetrization
\begin{equation}
\hat{{\rm R}}_{\mu\nu}=\hat{R}_{(\mu\nu)}.
\end{equation}
Now, we solve \eqref{NCein} for $g=g_s+h$ where $g_s$ is the Schwarzschild background
\begin{equation}
	ds^2=-f(r)dt^2+\frac{1}{f(r)}dr^2+r^2d\Omega^2,
\end{equation}
where $f(r)=1-R/r$, $R=2GM/c^2$ is the Schwarzschild radius and $h$ are axial modes \cite{proceeding} of perturbation that in the Regge-Wheeler gauge  have only the following non-zero components \cite{Roussille:2022vfa}
\begin{equation}\label{ansatz}
\begin{split}
	h_{t \theta}&=\frac{1}{\sin \theta} \sum_{\ell, m} h_0^{\ell m} \partial_{\phi} Y_{\ell m}(\theta, \phi)e^{-i \omega t},\\
	 h_{t \phi}&=-\sin \theta \sum_{\ell, m} h_0^{\ell m} \partial_\theta Y_{\ell m}(\theta, \phi)e^{-i \omega t},\\
	h_{r \theta}&=\frac{1}{\sin \theta} \sum_{\ell, m} h_1^{\ell m} \partial_{\phi} Y_{\ell m}(\theta, \phi) e^{-i \omega t},\\
	 h_{r \phi}&=-\sin \theta \sum_{\ell, m} h_1^{\ell m} \partial_\theta Y_{\ell m}(\theta, \phi) e^{-i \omega t}.
\end{split}
\end{equation}
where $h^{\ell m}_0$ and $h^{\ell m}_1$ are just functions of $r$. Surprisingly, as in the commutative case \cite{Regge:1957td, Edelstein:1970sk}, from the 10 equations in \eqref{NCein}, 3 are identically zero, while from the 7 remaining there are  only 3 that differ in the radial part. After factoring out the common angular and temporal part, from $\hat{{\rm R}}_{r\phi}=0$ we get
\begin{equation}
\begin{split}
       &  4 i r^4 (r - R)\omega h_0 + 2 r^2 (r - R) \big(r^3 \omega^2 - (r - R)(\ell(\ell + 1)\\
			&-2)\big)h_1 - 2 i \omega r^5(r - R) h_0' \\
 & + \qbar m \Big[ 2 i r^3 \omega (r - 2R) h_0 +
\big((2\ell(\ell+1)+12) r (r- R)^2 \\
&- 9(r - R)^2 R - r^4 R \omega^2\big)h_1 + i
r^4 R \omega h_0' + 2 r (r - R)^3 h_1' \Big] =0,
\end{split} 
\end{equation}
from $\hat{{\rm R}}_{t\phi}=0$ we get
\begin{equation}
\begin{split}
          &2r (2R - \ell(\ell+1)r) h_0 +
4 i r^2 \omega (r - R) h_1 \\
 & + 2 r^3(r - R)(i \omega h_1' + h_0'') + \qbar m \Big[  (2 \ell (\ell+1) r + R)h_0  \\
 & + i r \omega (4 r - 3 R) h_1 + r (4r - 5 R) h_0' + r^2 R ( i\omega h_1' + h_0'') \Big] = 0,
\end{split} 
\end{equation}
and from $\hat{{\rm R}}_{\theta\phi}=0$ we get
\begin{equation}
\begin{split}
     & \frac{i r^3 \omega}{r - R}h_0 + R h_1 + r(r
- R)h_1'  \\
     &- \qbar m \Big[ \frac{i r^2 R \omega}{2(r - R)^2}h_0 - 3
\frac{r - R}{r} h_1 - \frac{1}{2}R h_1'\Big]=0.
\end{split}
\end{equation}
Finally, from the remaining components, i.e. $\hat{{\rm R}}_{r \theta}=0$, $\hat{{\rm R}}_{\theta \theta}=0$, $\hat{{\rm R}}_{\varphi \varphi}=0$, and $\hat{{\rm R}}_{t \theta}=0$ we get equations that are identical to the above ones.

\begin{figure}[h]
\centering
\begin{tikzpicture}
\begin{axis}[
    axis lines = left,
    xlabel = \(r\),
    ylabel = {\(V(r)\)},
xmin=1,
    xmax=10,
    ymin=-0.72,
    ymax=0.3,
    legend cell align={left},
    legend style={at={(1,0)},anchor=south east}]
\addplot [
    domain=1:10, 
    samples=300, 
    color=black,
]
{(x-2)*(2*(2+1)*x-6)/x^4};
\addlegendentry{\tiny\(V_{RW}(r)\)}

\addplot [
    domain=1:10, 
    samples=300, 
    color=blue,
    ]
    {(x-2)*(2*(2+1)*x-6)/x^4+0.2*(2*(2+1)*(6-2*x)*x+2*(5*x-16))/(2*x^5)};
\addlegendentry{\tiny \(V(r), qm=0.2\)}

\addplot [
    domain=1:10, 
    samples=300, 
    color=red,
    ]
    {(x-2)*(2*(2+1)*x-6)/x^4-0.2*(2*(2+1)*(6-2*x)*x+2*(5*x-16))/(2*x^5)};
\addlegendentry{\tiny \(V(r), qm=-0.2\)}

\addplot [
    domain=1:10, 
    samples=300, 
    color=green,
    ]
    {(x-2)*(2*(2+1)*x-6)/x^4+0.1*(2*(2+1)*(6-2*x)*x+2*(5*x-16))/(2*x^5)};
\addlegendentry{\tiny \(V(r), qm=0.1\)}

\addplot [
    domain=1:10, 
    samples=300, 
    color=pink,
    ]
    {(x-2)*(2*(2+1)*x-6)/x^4-0.1*(2*(2+1)*(6-2*x)*x+2*(5*x-16))/(2*x^5)};
\addlegendentry{\tiny \(V(r), qm=-0.1\)}

\end{axis}
\end{tikzpicture}
\caption{The figure is drawn for $R=2$ ($M=1$) and $\ell=2$. We see that the behavior is very similar to the commutative case outside the horizon, and especially around the peak $r_0\cong 1.5 R$ and for larger $r$.} \label{proba}
\end{figure}
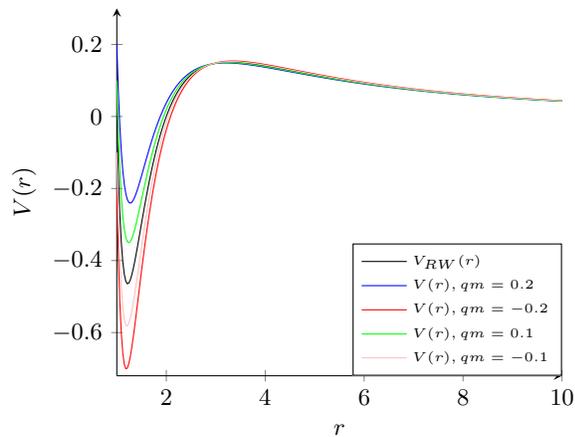

From  the 3 equations above only 2 are mutually independent, and after combining them one can get a single second order differential equation for $h_1$ (see \cite{novi} for a detailed proof)
\begin{equation}\begin{split}
	&r (r-R) \Big( \ell(\ell + 1) r (R-r) + 2r^2 - 6rR + 5R^2 + \omega^2 r^4 \Big) h_1 \\
	&+r^2(r - R)^2\Big( (5R - 2r)h_1' + r(r - R)h_1''\Big) + \\
         &\qbar m \bigg[\Big(\ell(\ell + 1)r(r - R)^2 - 6r^3 + \frac{R}{2}(49
r^2 - 64 r R + 26 R^2\\
& - \omega^2 r^4) \Big) h_1
 + r(r - R)^2 \Big( 3 ( r
- 2R) h_1' + \frac{1}{2}r R h_1''\Big) \bigg]=0,
\end{split}
\end{equation}
which after the redefinition of the field
\begin{equation}
h_1(r) = \frac{r^2}{r - R}\Big[ 1 + \frac{\qbar m}{2} \Big( \frac{3}{r} - \frac{1}{r - R} + \frac{1}{R} \log \frac{r}{r - R} \Big) \Big]\psi(r)
\end{equation}
and the change of variable\footnote{Notice that for negative values of $q m$ the variable $r_*$ behaves like in the commutative case.
It tends to $-\infty$ as $r \to R$. For positive values of $q m$ it still decreases, but at some point, around $q m$ away from the horizon, diverges to $+\infty$. This "pathology" can be regularized by appropriate change of variables so that $\hat{r}_* \to -\infty$ as $r \to R$ in both cases. Then boundary conditions can unambiguously be imposed and spectrum can be obtained by any method, including WKB.}
\begin{equation}
r_* = r + R \log \frac{r - R}{R} + \frac{\qbar m}{2} \frac{R}{r - R}
\end{equation}
reduces to the Schr\"{o}dinger form
\begin{equation}\label{qRW}
\frac{d^2 \psi}{d r_*^2} + \Big( \omega^2 - V(r) \Big) \psi = 0,
\end{equation}
where $V(r)=V_{RW}+V_{\qbar}$ is the $\qbar$uantum Regge-Wheeler potential given by
\begin{equation}\begin{split}\label{potential}
V(r) &= \frac{(r - R)\big(\ell (\ell + 1)r - 3R\big)}{r^4}\\
& + \qbar m \frac{\ell(\ell + 1)(3R - 2r)r + R(5r - 8R)}{2 r^5}.
\end{split}
\end{equation}
$V_{RW}$ is the usual Regge-Wheeler potential \cite{Regge:1957td, Edelstein:1970sk}, while $V_{\qbar}$ is the $\qbar$-correction. 
Unlike the standard Regge-Wheeler tortoise coordinate, the variable $r_*$ here defined is not a genuine coordinate on the manifold because it depends explicitly on $m$.
Notice that $V_{\qbar}$  depends on $\ell$ and $m$, so there will be a Zeeman-like effect where the potential of each $m$-mode is shifted as illustrated in FIG.\ref{proba}.

 In the commutative case the horizon $r=R$ is a zero of the potential $V_{RW}$. In the $\qbar$-deformed case we see that the condition $V(r)=0$ is fulfilled for a slightly shifted value close to the horizon, i.e. $r=R-\frac{\qbar m}{2} + \mathcal{O}(\qbar^2)$, which is also well illustrated in FIG.\ref{proba}. This means that each mode $m$ effectively ``sees'' a slightly different position of the horizon.

Also, it is interesting to note that the peak $r_0=1.5R$ of the potential $V_{RW}$ corresponds to the radius of the photon sphere, that is the last orbit of the photon. We see that in the $\qbar$-deformed case the peak also exhibits a minor shift (see FIG.\ref{proba11}) leading to a possibly interesting modification to the shadow of black hole \cite{Perlick:2021aok}.

\section{QNM frequencies}

For the QNM frequencies we need to see the asymptotic behavior of the $\qbar$uantum Regge-Wheeler equation \eqref{qRW} near the horizon and at the spatial infinity. It is easy to see that the near horizon behavior is modified with respect to the commutative theory, while the solution at spatial infinity remains the same. This will impact the QNM frequencies. There is a simple approach to calculate the QNM frequencies for systems described by equations of type \eqref{qRW} and potentials with a single peak \cite{Schutz:1985km}. 
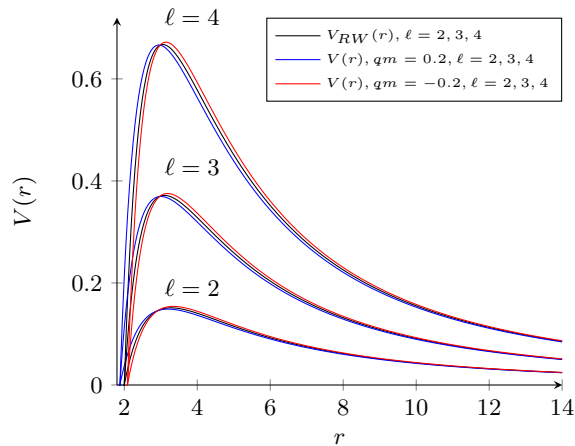
\begin{figure}[h]
\centering
\begin{tikzpicture}
\node[draw=none] at (1, 1.3) {$\ell = 2$};
\node[draw=none] at (1, 2.9) {$\ell = 3$};
\node[draw=none] at (1, 4.9) {$\ell = 4$};
\begin{axis}[
    axis lines = left,
    xlabel = \(r\),
    ylabel = {\(V(r)\)},
    xmin=1.8,
    xmax=14,
    ymin=0.00,
    ymax=0.72,
    legend cell align={left},
    legend style={at={(1,1)},anchor=north east}]
\addplot [
    domain=1.7:14, 
    samples=250, 
    color=black,
]
{(x-2)*(2*(2+1)*x-6)/x^4};
\addlegendentry{\tiny\(V_{RW}(r), \ell=2,3,4\)}
\addplot [
    domain=1.7:14, 
    samples=250, 
    color=blue,
    ]
    {(x-2)*(2*(2+1)*x-6)/x^4+0.2*(2*(2+1)*(6-2*x)*x+2*(5*x-16))/(2*x^5)};
\addlegendentry{\tiny \(V(r), \qbar m=0.2, \ell=2,3,4\)}
\addplot [
    domain=1.7:14, 
    samples=250, 
    color=red,
    ]
    {(x-2)*(2*(2+1)*x-6)/x^4-0.2*(2*(2+1)*(6-2*x)*x+2*(5*x-16))/(2*x^5)};
\addlegendentry{\tiny \(V(r), \qbar m=-0.2, \ell=2,3,4\)}

\addplot [
    domain=1.7:14, 
    samples=250, 
    color=black,
]
{(x-2)*(3*(3+1)*x-6)/x^4};

\addplot [
    domain=1.7:14, 
    samples=250, 
    color=black,
]
{(x-2)*(4*(4+1)*x-6)/x^4};

\addplot [
    domain=1.7:14, 
    samples=250, 
    color=blue,
    ]
    {(x-2)*(3*(3+1)*x-6)/x^4+0.2*(3*(3+1)*(6-2*x)*x+2*(5*x-16))/(2*x^5)};
\addplot [
    domain=1.7:14, 
    samples=250, 
    color=blue,
    ]
    {(x-2)*(4*(4+1)*x-6)/x^4+0.2*(4*(4+1)*(6-2*x)*x+2*(5*x-16))/(2*x^5)};
\addplot [
    domain=1.7:14, 
    samples=250, 
    color=red,
    ]
    {(x-2)*(3*(3+1)*x-6)/x^4-0.2*(3*(3+1)*(6-2*x)*x+2*(5*x-16))/(2*x^5)};

\addplot [
    domain=1.7:14, 
    samples=250, 
    color=red,
    ]
    {(x-2)*(4*(4+1)*x-6)/x^4-0.2*(4*(4+1)*(6-2*x)*x+2*(5*x-16))/(2*x^5)};
\end{axis}
\end{tikzpicture}
\caption{The figure is drawn for $R=2$ ($M=1$) and $\ell=2,3,4$ around the peak.} \label{proba11}
\end{figure}
This is a semi-analytical method based on the usual WKB approximation where one looks at wave scattering on the peak of the potential barrier (as seen on FIG.\ref{proba11}). One should keep in mind that precision of the WKB approximation in this context is good enough to study qualitative aspects of the perturbation, but exact values of herein obtained frequencies should not be taken as definitive.
The complex QNM frequencies can be estimated using the third order WKB formulas given in \cite{Iyer:1986np}.
We give the QNM frequencies for the fundamental mode $n=0$ and for the orbital numbers  $\ell=2,3,4$ for various $\qbar m$ in FIG.\ref{tab1}.
\begin{figure}[h]
\centering
\begin{center}
\begin{tabular}{rccccc}
$\qbar m = $ & $-0.2$ & $-0.1$ & $0$ & $0.1$ & $0.2$ \\
\\
$\ell = 2$ & & & & & \\
\hline
$\mathrm{Re}(\omega_0)$ & $ \ 0.37712 \ $ & $ 0.37507 \ $ & $ 0.37316 \ $ & $ 0.37144  \ $ & $ 0.36991 \ $ \\
$-\mathrm{Im}(\omega_0)$ & $ \ 0.08873 \ $ & $ 0.08906 \ $ & $ 0.08922 \ $ & $ 0.08898 \ $ & $ 0.08782 \ $ \\
 & & & & & \\
$\ell = 3$ & & & & & \\
\hline
$\mathrm{Re}(\omega_0)$ & $0.60260$ & $0.60076$ & $0.59927$ & $0.59827$ & $0.59800$ \\
$-\mathrm{Im}(\omega_0)$ & $0.09243$ & $0.09265$ & $ 0.09273 $ & $0.09246$ & $0.09132$ \\
 & & & & & \\
$\ell = 4$ & & & & & \\
\hline
$\mathrm{Re}(\omega_0)$ & $0.81218$ & $0.81038$ & $0.80910$ & $0.80852$ & $0.80897$ \\
$-\mathrm{Im}(\omega_0)$ & $0.09390$ & $0.09410$ & $0.09417$ & $0.09397$ & $0.09310$ \\
\end{tabular}
\end{center}
\caption{The table of QNM's for $n=0$, $M=1$, $\ell=2,3,4$.} \label{tab1}
\end{figure}
From the FIG.\ref{tab1} we see that the imaginary part of the QNM frequency is always negative which leads to the conclusion that the NC gravity theory is also stable under linear perturbations.

It is a well known result, both using WKB \cite{Schutz:1985km} and numerical analysis \cite{Chandrasekhar:1975zza} that the imaginary part of QNM  frequency $\omega_0$ saturates to $-0.096225$ in the limit $\ell\longrightarrow\infty$. We observe that this limiting property also exists in the $\qbar$-deformed case, and the  values of the imaginary part of QNM frequency remain negative. The saturated values are $-0.095997$ and $-0.095914$ for $\qbar m = -0.2$ and $\qbar m = 0.2$ respectively.

\section{Conclusion}
\label{SECconclusion}

In this paper we have presented a formulation of the Regge-Wheeler equation in a NC framework, which is a candidate for describing the quantum structure of the spacetime. We used a mathematically well founded formalism of generalized symmetries (Hopf algebras) and the NC differential geometry for defining NC geometric objects (like curvature, torsion etc.). The choice of twist \eqref{qtwist} is argued in section II and is due to several reasons: symmetry, non-triviality and simplicity. Namely, the most general NC spacetime algebra would be described by 
\[  
[\hat{x}_{\mu}, \hat{x}_{\nu}]=\hat{\Theta}_{\mu\nu}(\hat{x})=\Theta_{\mu\nu}+C_{\mu\nu}^{\alpha}\hat{x}_{\alpha}+D_{\mu\nu}^{\alpha\beta}\hat{x}_{\alpha}\hat{x}_{\beta}+\mathcal{O}(\hat{x}^3)
 \]
which in the low energy regime is determined by the constant tensor $\Theta$, i.e. the Moyal space, for which we can use the NC differential geometry framework based on the Moyal twist \eqref{twist}.  The semi-Killing form of twist is chosen because it gives non-trivial corrections to the curvature tensor when regarding perturbations of Schwarzschild solution (the Killing one has no corrections!), and its form produces the ``simple equations'' like \eqref{RiT} due to nice form in spherical coordinates that in the end lead to a separable set of differential equations (in the standard RW gauge) culminating in (20-23). One can work with more general vector fields but then equations like \eqref{RiT} would get extra contributions and the whole calculation would be much more involved, meaning that the separability issue became serious technical challenge.
The question of symmetries was also considered while constructing the twist \eqref{qtwist}. Namely, symmetry is something we usually impose (given some hint from phenomenology). Even in the commutative case the Einstein field equations can be solved only once some symmetry is imposed. We expect similar behavior for NC case, that is we need NC field equation and some (NC) symmetry. In our case the symmetry is governed by the unperturbed solution, namely the Schwarzschild background, which was shown to provide nontrivial NC corrections only if a portion of these symmetries is kept in the deformation procedure via twist.  In a sense, the twist \eqref{qtwist} is \emph{a posteriori} chosen once all the above was taken into consideration.

The further analysis in this paper has been done in the lowest order of the NC parameter $\qbar$. Up to that order, the potential term $V(r)$ in the $\qbar$RW equation \eqref{qRW} describing the QNM's bifurcates into a commutative part $V_{RW}$ and another part proportional to the NC parameter $V_{\qbar}$. Thus, in the limit of vanishing NC parameter, we recover the conventional Regge-Wheeler equation in the commutative spacetime.

The time dependence of the QNM's is governed by the term $ e^{-i \omega t}$ , which is explicit in \eqref{ansatz}. It is well known that in the commutative case, the Schwarzschild black hole is stable under linear gravitational perturbations, which is indicated by the fact that the imaginary part of the QNM frequency is negative \cite{natureV, visv}. As it can be seen from FIG. 3  the imaginary parts of the QNM's in the presence of noncommutativity remain negative. This indicates that the Schwarzschild black hole remains stable under the linear gravitational perturbations even in the NC framework. 
This is suggestive of the modified stability of black holes at the Planck scale, which may have consequences for the existence of the primordial black holes. A qualitatively similar results were obtained using  Loop Quantum Gravity inspired black holes and the corresponding quantum Regge-Wheeler potential \cite{Cruz:2015bcj}. 

The numerical values of QNM frequencies in FIG.\ref{tab1}, together with the potentials illustrated in FIG.\ref{proba} and \ref{proba11} are calculated in the  $-0.2\leq \qbar m \leq 0.2$ range. These values should be taken with a grain of salt as $\qbar m = 0.2$ in natural units corresponds to $0.2M$, i.e. $0.2R/2=0.1R$, that is $\qbar$ is around $10\%$ of the black hole radius $R$. We chose this range so we could illustrate the general qualitative behavior of the NC effects, namely the Zeeman-like effect for the potential and the persistence of the stability in the NC framework. It is more realistic that this range is somewhere closer to the Planck scale. Namely there is a series of papers \cite{Calmet:2004dn, Carroll:2001ws, Zhang:2004yu, Piscicchia:2022xra, Joby:2014oee, Kobakhidze:2016cqh, Jenks:NCG} where various bounds are put on the NC deformation parameter, with $\qbar\approx 10^{-34}{\rm m}$ being the strongest one. 
For these kind of small values for $\qbar$ we need to go to a much higher precision in calculating QNM frequencies. 
One can use direct integration method \cite{Gundlach:1993tp} or Leaver method \cite{Leaver:1985ax, Leaver:1990zz, Nollert:1993zz}.

The presentation in this paper is adapted for the Schwarzschild black hole, but the same formalism can be generalized to other black holes including Kerr, as well as other compact objects such as neutron stars and pulsars, which offers the possibility of examining the NANOGrav data for the gravitational waves from pulsars \cite{nano, clanak} within the NC context. In addition, our formalism can be used to study a much wider class of NC spaces and the associated NC parameters can provide a parametrized form of the predicted QNM spectrum which can be confronted with data from future detectors such as LISA.\\

\section*{Acknowledgments}

This  research was supported by the Croatian Science
Foundation Project No. IP-2020-02-9614 {\it{Search for Quantum spacetime in Black Hole QNM spectrum and Gamma Ray Bursts.}}

\end{document}